\documentclass[preprint,tightenlines,aps,eqsecnum]{revtex4}
\usepackage{graphicx}

\begin{document}

\preprint{cond-mat/0308024}

\title{Competing orders in thermally fluctuating superconductors\\ in two dimensions}

\author{Subir Sachdev}
\affiliation{Department of Physics, Yale University, P.O. Box
208120, New Haven CT 06520-8120}

\author{Eugene Demler}
\affiliation{Department of Physics, Harvard University, Cambridge
MA 02138}

\date{August 1, 2003}

\begin{abstract}
We extend recent low temperature analyses of competing orders in
the cuprate superconductors to the pseudogap regime where all
orders are fluctuating. A universal continuum limit of a classical
Ginzburg-Landau functional is used to characterize fluctuations of
the superconducting order: this describes the crossover from
Gaussian fluctuations at high temperatures to the vortex-binding
physics near the onset of global phase coherence. These
fluctuations induce affiliated corrections in the correlations of
other orders, and in particular, in the different realizations of
charge order. Implications for scanning tunnelling spectroscopy
and neutron scattering experiments are noted: there may be a
regime of temperatures near the onset of superconductivity where
the charge order is enhanced with increasing temperatures.
\end{abstract}

\maketitle

\section{Introduction}
\label{sec:intro}

A number of recent perspectives
\cite{zaanennat,pnas,science,so5,sachdev_zhang,srmp,krmp} have
highlighted new experimental
\cite{lake,halperin,boris,lake2,seamus,howald} and theoretical
\cite{kwon,prl,vadim,tolyastm,vojta,zaanen,podolsky,chenting,zhu,franz,zhang,ghosal,andersen02}
works exploring the interplay between the multiple order
parameters which characterize the ground state of some of the
cuprate superconductors. Good evidence was obtained for a strong
coupling between the superconducting order and density wave order
in spin/charge/bond correlations (described more precisely below).
In particular, by tuning the superconducting order by an applied
magnetic field at very low temperatures ($T$), a strong
field-dependent variation was observed in the latter correlations.

In this paper, we explore the possibility of observing related
connections in the finite temperature `pseudogap' region above the
superconducting critical temperature, $T_c$. Here, the
superconducting order has strong $T$ dependent fluctuations; we
will compute these fluctuations in the framework of a
two-dimensional Ginzburg-Landau theory, including a precise
characterization of strong fluctuations obtained from numerical
studies. We will show that the model of Ref.~\onlinecite{prl}
predicts that such fluctuations lead to a corresponding
sympathetic variation in the autocorrelations of the other orders.
Working to linear order in the coupling between superconductivity
and these orders, we provide a computation of certain universal
characteristics of the $T$ dependence of the latter fluctuations.
Our results will also be formally extended to $T < T_c$ for
completeness, but it must be noted that we neglect the inter-layer
coupling and quantum effects, which become important at lower $T$
\cite{3d}.

We begin by defining the order parameters under consideration. The
primary order is the complex superconducting order $\Psi ({\bf
r})$ which describes the spatial variation in the order associated
with condensation of Cooper pairs. This is expected to undergo
strong `phase' fluctuations \cite{phase} for $T$ near $T_c$. Using
the proximity of the underdoped cuprates to a superfluid-insulator
quantum transition, Refs.~\onlinecite{ssrelax,nature} argued that
`amplitude' fluctuations should be treated at an equal footing
\cite{gap}, and proposed that such thermal fluctuations could be
described by a classical partition function of a suitable
universal continuum limit of the Ginzburg-Landau free energy: this
will be reviewed here in Section~\ref{sec:sim}. Such an approach
describes the crossover from Gaussian superconducting fluctuations
at temperatures well above $T_c$, to the vortex physics of the
Kosterlitz-Thouless transition near $T_c$. A dynamic theory with a
similar static component (although with a lattice cutoff) was
recently used \cite{huse1,ussi,huse2} to describe the notable
measurements \cite{ong} of the Nernst effect.

This fluctuating superconductor is also expected to have
appreciable correlations in other order parameters. The
spin-density-wave order is described by the complex 3-component
vectors $\Phi_{x\alpha}$, $\Phi_{y\alpha}$, where
$\alpha=x,y,z$ extends over the 3 spin directions, and the spin
operator on site ${\bf r}$, $S_{\alpha} ({\bf r})$ is given by
\begin{equation}
S_{\alpha} ({\bf r}) = \mbox{Re} \left[e^{i {\bf K}_{sx} \cdot
{\bf r}} \Phi_{x \alpha} ({\bf r})+  e^{i {\bf K}_{sy} \cdot {\bf
r}} \Phi_{y \alpha}({\bf r})\right]. \label{eq1}
\end{equation}
Here ${\bf K}_{sx,y}$ are the spin-density-wave ordering
wavevectors along the $x$ and $y$ principle axes of the square
lattice: near a doping of $\delta=1/8$, we have ${\bf
K}_{sx}=(3\pi/4,\pi)$ and ${\bf K}_{sy}=(\pi,3\pi/4)$. In a
similar manner we can define  bond order parameters $\phi_{{\bf
a}x,y} ({\bf r})$ by examining the modulations in the exchange
energy of a pair of spins separated by a distance ${\bf a}$:
\begin{equation}
S_{\alpha} ({\bf r}) S_{\alpha} ({\bf r} + {\bf a}) =  \mbox{Re}
\left[ e^{ i {\bf K}_{cx} \cdot {\bf r}} \phi_{{\bf a}x} ({\bf r})
+ e^{ i {\bf K}_{cy} \cdot {\bf r}} \phi_{{\bf a}y} ({\bf
r})\right]. \label{eq2}
\end{equation}
The special case ${\bf a}=0$ of $\phi_{{\bf a}x,y}$ is a measure
of the charge density wave order. Comparison between (\ref{eq2})
and (\ref{eq1}) suggests that the ordering wavevectors are related
by ${\bf K}_{cx,y} = 2 {\bf K}_{sx,y}$, and this is observed
experimentally.

A number of other order parameters which are invariant under spin
rotations, like $\phi_{{\bf a}x,y}$, can also be defined
\cite{prl,podolsky}. These include the site charge density, the
average electron kinetic energy in a bond, or modulations in the
pairing amplitude. By symmetry, all such quantities will have
modulations at the wavevectors ${\bf K}_{cx,y}$, and we can
therefore expect that their order parameter fluctuations will
track those of $\phi_{{\bf a}x,y}$. Differences in microscopic
physics can, of course, make some of these modulations much larger
than others. We will not explicitly consider all such
possibilities here, and the reader should view $\phi_{{\bf a}x,y}$
as a suitable representative of the order parameters
characterizing modulations at the wavevectors ${\bf K}_{cx,y}$ in
observables invariant under spin rotations. We will subsequently
refer to the order represented by $\phi$ simply as charge order.

While the focus of this paper is on the interplay between the
superconducting fluctuations and the orders mentioned above, it
should be clear to the reader that our considerations are quite
general. Simple extensions lead to similar effects in the
interplay of superconductivity with most of the other orders in
the zoo of possibilities considered in the theory of the cuprates.

In considering correlations of $\Psi$, $\Phi$ and $\phi$ in the
fluctuation region, it is important to consider the influence of
random static impurities which are invariably present in the
cuprates. As almost all impurities preserve electron number and
spin rotation invariance, their influence on $\Psi$ and $\Phi$
will consist of perturbations in the {\em random exchange} class
(this is discussed more explicitly in Section~\ref{sec:corr}). In
contrast, the order $\phi$ breaks only lattice symmetries, and is
consequently subject to the far more disruptive {\em random field}
perturbations \cite{apy}. In two spatial dimensions, this implies
that true long-range order cannot develop as $T \rightarrow 0$,
and that the $\phi$ correlation length saturates at a finite
value. We will assume here that there is a local onset of $\phi$,
$\Phi$, and $\Psi$ orders at temperatures where the pseudogap
develops, but at lower temperatures $\phi$ correlations are
predominantly controlled by the random-field disorder, and have
only a weak, intrinsic $T$ dependence. This is also consonant with
the result that thermal fluctuations are irrelevant at the
random-field transition in higher dimensions \cite{apy}. In
contrast, the fluctuations of $\Psi$ and $\Phi$ are strongly $T$
dependent, and can have an infinite correlation length as $T
\rightarrow 0$. The $\Psi$ order becomes quasi-long-ranged at
$T=T_c$ and has the strongly $T$-dependent Gaussian-to-vortex
crossover noted above at $T>T_c$. The $\Phi$ order can also have
the exponential rapid $T$ dependence associated with the breaking
of O(3) spin rotation symmetry as $T \rightarrow 0$.

This paper will consider the regime above $T_c$ where
\begin{equation}
\langle \Psi ({\bf r}) \rangle = 0~~~;~~~\langle \Phi_{x,y\alpha}
({\bf r})\rangle = 0~~~;~~~\langle \phi_{{\bf a}x,y} ({\bf r})
\rangle \neq 0 \label{eq3}
\end{equation}
The non-zero value $\langle \phi \rangle$ is due to the presence
of random-field perturbations which explicitly break lattice
symmetries, and so allow $\phi$ to locally have a non-zero mean
value which will fluctuate randomly as a function of ${\bf r}$. As
noted above, we assume that $\langle \phi \rangle$ only has a weak
intrinsic $T$ dependence. However, the fluctuations of the $\Psi$,
$\Phi$, and $\phi$ orders are not independent, and so the strong
$T$ dependence associated with the Gaussian-to-vortex crossover in
$\Psi$ will induce a corresponding $T$-dependent variation in
$\langle \phi \rangle$. This paper will compute this variation and
suggest associated experimental tests. Strictly speaking, because
there is only quasi-long-range order in $\Psi$ below $T_c$, the
expectation values (\ref{eq3}) apply also for $T<T_c$: indeed, our
methods and results extend also to $T<T_c$. However, as noted
earlier, we neglect the effects of inter-layer couplings and of
quantum fluctuations, and so our low $T$ results should be treated
with caution.

Our theory for the fluctuating orders and their interplay is
summarized in Section~\ref{sec:corr}, which also contains are main
results. Details of the continuum theory of the superconducting
fluctuations and its Gaussian-to-vortex crossover appear in
Section~\ref{sec:sim}. Section~\ref{sec:conc} discusses
experimental tests and possible extensions of our theory.

\section{Correlations between fluctuating orders: main results}
\label{sec:corr}

This section will introduce the free energies which control the
fluctuations of the order parameters, and state our main results
on the $T$ dependence of the $\phi$ order at $T>T_c$.

We describe the fluctuations of the superconducting order $\Psi
({\bf r})$ by a classical continuum partition function over the
Ginzburg-Landau free energy \cite{ssrelax}
\begin{eqnarray}
\mathcal{Z}_{GL} &=& \int \mathcal{D} \Psi ({\bf r})
e^{-\mathcal{F}_{GL}/(k_B T)} \nonumber \\
 \mathcal{F}_{GL} &=& \int
d^2 r \left[ \frac{\hbar^2}{2 m^{\ast}} |\nabla_{{\bf r}} \Psi
({\bf r})|^2 + a(T) |\Psi ({\bf r})|^2 + \frac{b}{2} |\Psi ({\bf
r}) |^4 \right] \label{gl}
\end{eqnarray}
We use here the notation of Refs~\onlinecite{huse1,ussi,huse2}:
$m^{\ast}$, $a(T)$, $b$ are parameters which can be computed, in
principle, from the microscopic physics of the underlying
electrons. The co-efficient of $|\Psi ({\bf r})|^2$, $a(T)$,
vanishes at a mean-field transition temperature, $a(T_c^{MF})=0$,
which will be distinct from the temperature $T_c$ at which there
is a Kosterlitz-Thouless transition {\em i.e.} $a(T_c) < 0$. The
purely two-dimensional, and classical theory (\ref{gl}) is
expected to apply to the cuprates only for $T>T_c$: below $T_c$ we
have to also account for three-dimensional effects arising from
inter-layer couplings, and for quantum effects at low enough $T$.
All such effects will be neglected here, but for completeness, we
will nevertheless discuss properties of the theory (\ref{gl}) over
the full range of $T$ values.

An important point is that the functional integral in (\ref{gl})
is not defined on its own, and needs an ultraviolet regulator. In
the physical system this is provided by the underlying electron
physics on the lattice, but this is very difficult to characterize
explicitly. Here, we shall follow the procedure proposed in
Ref.~\onlinecite{ssrelax}: the ultraviolet dependence can be
accounted for by a suitable renormalization in the value of
$a(T)$. However, because we do not know the explicit form of the
ultraviolet cutoff, we cannot {\em a priori} compute the needed
shift in $a(T)$. This lack of knowledge can be circumvented by
using the experimental value of $T_c$ as an input into our
calculation. The knowledge of the actual $T_c$, combined with the
parameters in (\ref{gl}) then allows a quantitative computation of
the Gaussian-to-vortex crossover with no free parameters. We
re-iterate that (\ref{gl}) cannot be regarded as a fully
predictive theory on its own, and so cannot, even in principle,
predict the actual value of $T_c$: once $T_c$ is determined by
other means, precise quantitative predictions for other
observables become possible.

The Gaussian-to-vortex crossover can be expressed in terms of the
following dimensionless parameter
\begin{equation}
g \equiv \frac{\hbar^2}{m^{\ast} b} \left[ \frac{a(T)}{k_B T} -
\frac{a(T_c)}{k_B T_c} \right]. \label{defg}
\end{equation}
The parameter $g$ should be a monotonically increasing function of
$T$. For $T \ll T_c$, $g \sim -1/T$, at $T=T_c$, we have $g=0$,
and above $T_c$, $g$ takes positive values. We will see later that
the present continuum theory eventually breaks down at large $T$,
when $g$ begins to acquire a non-monotonic dependence on $T$. The
value of $1/|g|$ is a measure of the strength of corrections to
the mean field theory of $\mathcal{Z}_{GL}$.

It is important to note that the $T$ dependence of $a(T)$ in
(\ref{gl}) and (\ref{defg}) is non-universal, and this will lead
to some non-universality in the $T$ dependence of all our
predictions. However, one of our main points is that there is a
universal dependence on the parameter $g$. Moreover, once we
assume a linear $T$ dependence of $a(T)$ near $T_c$ (as is
commonly done, and we will do in (\ref{mh})), the $T$ dependence
of our predictions becomes specific.

Aided by the results of
Ref.~\onlinecite{ssrelax,nikolay1,nikolay2} we will show that it
is possible to obtain precise predictions for a variety of
correlators of $\mathcal{Z}_{GL}$. We quote a result which will be
useful in our analysis here of multiple order parameters:
\begin{equation}
\frac{\hbar^2}{m^{\ast}} \left[\frac{\left\langle \left| \Psi
\right|^2 \right\rangle_{T}}{k_B T} - \frac{\left\langle \left|
\Psi \right|^2 \right\rangle_{T_c}}{k_B T_c} \right] =
\mathcal{D}(g, T/T_c) \label{asymp1}
\end{equation}
where $\mathcal{D}(g, T/T_c)$ is a universal function. The
averages on the left-hand-side are evaluated under the partition
function $\mathcal{Z}_{GL}$ at the indicated temperature. We will
show in Section~\ref{sec:sim} that it is possible to re-express
the two argument function $\mathcal{D} (g, T/T_c)$ in terms of a
single argument function $F(\mathcal{G})$ as in (\ref{final}),
where $\mathcal{G}$ depends upon $g$ and $T/T_c$ as in
(\ref{defgG}). Here we present results for the initial crossover
from the Gaussian to the vortex regime, which occurs when $g \gg
1$:
\begin{eqnarray}
\mathcal{D} (g, T/T_c) &=& - \frac{1}{2 \pi} \ln (380 g T/T_c) +
\frac{1}{2
\pi^2 g} \ln (13.3 g T/T_c) \nonumber \\
&~&~~~~~~~~~~~+ \frac{1}{4 \pi^3 g^2} \left[ \ln^2 (13.3g T/T_c) -
2\ln (7.86 g T/T_c) \right]+ \mathcal{O}(1/g^3). \label{asymp2}
\end{eqnarray}
The numerical constants appearing in the arguments of the
logarithms are universal. These constants, and the constants
appearing in the arguments of all subsequent logarithms, depend on
only two universal numbers that have to be determined by computer
simulations: the latter numbers are the constant $\mathcal{G}_c$
computed first in Ref.~\onlinecite{ssrelax}, and the constant
$\xi$ computed in Refs.~\onlinecite{nikolay1,nikolay2}. Additional
higher order terms in (\ref{asymp2}) have also been computed and
these will be presented in Section~\ref{sec:sim}: we show there
that it is possible to account for {\em all} logarithmic terms
that appear at higher orders in $g$. Numerical results for the
full range of values of $g$ appear in Section~\ref{sec:sim}. The
expression (\ref{asymp1}) has ignored the possible
$T$-dependencies of $m^{\ast}$ and $b$ for simplicity: it is
possible to account for these in a similar manner, as will become
clear from the discussion in Section~\ref{sec:sim}.

It is worth noting here that vortices are already present in the
Gaussian theory, associated with zeros of $\Psi ({\bf r})$
\cite{bert}. The result (\ref{asymp2}) accounts for the initial
correlations between these vortices, but does not include the
vortex-binding physics of the Kosterlitz-Thouless transition. The
latter is only accounted for by the numerical results in
Section~\ref{sec:sim}.

Let us turn now to the density wave order parameters $\Phi$,
$\phi$. The complete effective action for these order parameters
has a rather complicated structure and was discussed in
Ref.~\onlinecite{prl}. A simple Gaussian form will be satisfactory
for our purposes here:
\begin{eqnarray}
\mathcal{F}_{\Phi} &=& \int d^2 r\left[ K_{\Phi x} |\nabla_x
\Phi_{x \alpha} |^2 + K_{\Phi y} |\nabla_y \Phi_{x \alpha} |^2 +
\xi_{\Phi}^{-2} |\Phi_{x \alpha} |^2 \right. \nonumber \\
&~&~~~~~~~~~~~~~~~~~\left. + h_{\Phi x} ({\bf r}) \Phi_{\alpha}^2
({\bf r})+ h_{\Phi x}^{\ast} ({\bf r}) \Phi_{\alpha}^{\ast 2}
({\bf r}) + (x \leftrightarrow y) +
  \ldots \right] \nonumber \\
\mathcal{F}_{\phi} &=& \int d^2 r\left[ K_{\phi x} |\nabla_x
\phi_{{\bf a} x} |^2 + K_{\phi y} |\nabla_y \phi_{{\bf a} x} |^2 +
\xi_{\phi}^{-2} |\phi_{{\bf a} x} |^2 \right. \nonumber \\
&~&~~~~~~~~~~~~~~~~~\left.+ h_{\phi x} ({\bf r}) \phi_{{\bf a}x}
({\bf r}) + h_{\phi x}^{\ast} ({\bf r}) \phi_{{\bf a} x} ({\bf r})
+ (x \leftrightarrow y) + \ldots \right] \label{FPp}
\end{eqnarray}
Apart from the usual Gaussian terms \cite{prl}, the above contains
complex random fields $h_{\Phi} ({\bf r})$ and $h_{\phi} ({\bf
r})$ which pin the `sliding' mode of the charge density wave.
These fields arise from impurities which preserve spin rotation
invariance: as a consequence, notice that the random coupling is
{\em linear} in the fields $\phi$, but that there is only a
random-exchange coupling to O(3) rotations in the spin density
wave order. These simple facts have a number of interesting
implications: \\
({\em i\/}) There can be no long range charge order in two spatial
dimensions, even at $T=0$. This implies that there can be no $T=0$
quantum critical point, tuned by the hole concentration,
associated with the onset of such order. A quantum critical point
associated with the restoration of O(3) symmetry
remains possible.\\
({\em ii\/}) The strong relevance of such random-field
perturbations suggests that in the absence of couplings to other
critical order parameters, the correlation length $\xi_{\phi}$
can be assumed to be roughly temperature-independent at low temperatures.\\
({\em iii\/}) The theories (\ref{gl}) and (\ref{FPp}) describe a
phase in which the expectation values in (\ref{eq3}) hold.

Finally, as promised, let us consider the influence of the $\Psi$
fluctuations described by $\mathcal{Z}_{GL}$ on the charge order
correlations. The simplest coupling between the orders is a
$\lambda |\Psi|^2 (|\phi_{{\bf a}x}|^2 + |\phi_{{\bf a}y}|^2 )$
term, and, as in Ref.~\onlinecite{prl}, this leads to the leading
order correction
\begin{equation}
\xi_{\phi}^{-2} (T) = \xi_{\phi 0}^{-2} (T) + \lambda
\left\langle \left| \Psi \right|^2 \right\rangle_{T}.
\label{eq10}
\end{equation}
Here $\xi_{\phi 0} (T)$ is the `bare' correlation length of the
$\phi$ order, which is expected to be only temperature dependent
near $T_c$. We input the value of $\langle |\Psi |^2 \rangle$ as
computed in (\ref{asymp1}) and Section~\ref{sec:sim}, and obtain
our main predictions for the superconducting fluctuation-induced
modification in the $\phi$ correlation length.

\section{Continuum theory of thermal superconducting fluctuations}
\label{sec:sim}

This section will review the results of Ref~\onlinecite{ssrelax}
relevant to obtaining (\ref{asymp1}) and (\ref{asymp2}) and its
extensions. Appendix~\ref{bosegas} will review the work of
Prokof'ev, Ruebenacker, and Svistinov \cite{nikolay1,nikolay2} on
the dilute two-dimensional Bose gas and show that the results of
their numerical simulations can be mapped onto universal
quantities needed here.

Ref.~\onlinecite{ssrelax} studied the following continuum theory
of a $N=2$ component real scalar $\varphi_{a}$, $a=1,2$:
\begin{equation}
\mathcal{F}_{\varphi} = \int d^2 r \left[ \frac{1}{2} \left(
\nabla_{{\bf r}} \varphi_a \right)^2 + \frac{\widetilde{R}}{2}
\varphi_a^2 + \frac{U}{24} \left( \varphi_a^2 \right)^2 \right]
\label{fphi}
\end{equation}
(Here we have changed notation for the field, from $\Phi_{\alpha}$
in Ref.~\onlinecite{ssrelax}, to $\varphi_a$ here, to prevent
confusion with the spin density wave order.) This theory maps onto
(\ref{gl}) with the following correspondences:
\begin{eqnarray}
\Psi &=& \sqrt{m^{\ast}} (\varphi_1 + i \varphi_2 )/\hbar
\nonumber \\
\widetilde{R} &=& 2 m^{\ast} a(T)/\hbar^2 \nonumber \\
U &=& 12 m^{\ast 2} b/\hbar^4 \label{defall}
\end{eqnarray}

It was argued \cite{ssrelax} that the continuum limit of
$\mathcal{F}_{\varphi}$ required only the single renormalization
of $\widetilde{R}$ to $R$:
\begin{equation}
\widetilde{R} = R - \frac{2k_B T U}{3}  \int^{\Lambda} \frac{d^2
k}{4 \pi^2} \frac{1}{k^2 + R}. \label{defR}
\end{equation}
Here we have introduced an ultraviolet cutoff $\Lambda$ which is
needed to regulate the theory $\mathcal{F}_{\varphi}$. The
renormalization in (\ref{defR}) is associated with logarithmic
ultraviolet divergence of the one-loop `tadpole' diagram; the
renormalized $R$ in the propagator on the right-hand-side accounts
for tadpoles-on-tadpoles etc. All other diagrams are ultraviolet
convergent, and hence the simple structure of the renormalization
theory.

It is important to note that (\ref{defR}) is the exact definition
of $R$, and consequently $R$ is {\em not} the fully `self-energy'
of the $\varphi_a$ field at zero external momentum; $R$ only
accounts for the resummation of tadpole graphs. In practice, the
relationship (\ref{defR}) implies that, when we perform a Feynman
graph expansion of any observable, we can ignore all tadpole
graphs, and replace $\widetilde{R}$ by $R$ in all propagators.
Notice also that as the bare coupling $\widetilde{R}$ extends from
$-\infty$ to $\infty$, the renormalized coupling $R$ extends from
0 to $\infty$.

After the renormalization of $\widetilde{R}$ to $R$, all
subsequent correlators of $\mathcal{F}_{\varphi}$ are ultra-violet
convergent, and so we can safely take $\Lambda \rightarrow \infty$
in them. This implies that all these correlators  are universal
functions of the single dimensionless quantity that can be
obtained from the parameters in (\ref{fphi}): this is the analog
of the `Ginzburg ratio', defined here as
\begin{equation}
\mathcal{G} = \frac{k_B T U}{R} \label{defG}
\end{equation}
For $T \ll T_c$, where $\widetilde{R} \ll 0$, we have $\mathcal{G}
\rightarrow \infty$. Conversely, for $T \gg T_c$, $\widetilde{R}
\gg 0$ and $\mathcal{G} \rightarrow 0$.

The field theory (\ref{fphi}) exhibits a Kosterlitz-Thouless
transition at some critical temperature, and the arguments above
imply that this transition occurs at a {\em universal} critical
value $\mathcal{G} = \mathcal{G}_c$. The numerical studies of
Ref.~\onlinecite{ssrelax} found $\mathcal{G}_c \approx 102$. The
value of $\mathcal{G}_c$ can also be obtained from the subsequent,
and more precise, numerical simulations of
Refs.~\onlinecite{nikolay1,nikolay2}; this connection is discussed
in Appendix~\ref{bosegas}, and (\ref{ximG}) yields
\begin{equation}
\mathcal{G}_c = 96.9 \pm 3\label{Gc}
\end{equation}

\begin{figure}
\centerline{\includegraphics[width=4.0in]{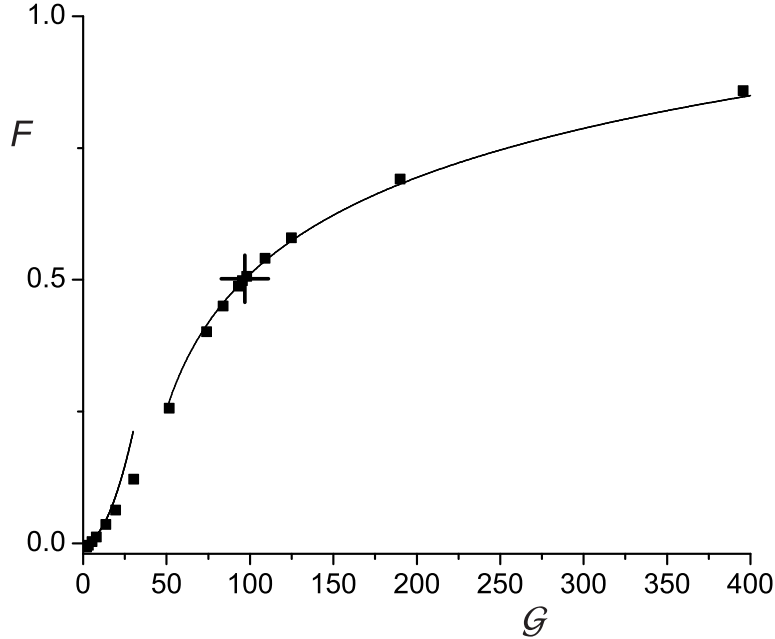}}
\caption{Plots of the universal function $F (\mathcal{G})$. The
line on the left is the small $\mathcal{G}$ approximation in
(\ref{smallG}). The line on the right is the large $\mathcal{G}$
approximation in (\ref{defgd}) and (\ref{largeG}). The square
symbols are the numerical data of
Ref.~\protect\onlinecite{nikolay2} transformed by (\ref{lambda})
and (\ref{defX}). The plus marks the position of the Kosterlitz
Thouless transition.} \label{plotF}
\end{figure}

We are interested here in the value of $\langle \varphi_a^2
\rangle$. This quantity requires a single additive renormalization
before the continuum limit is obtained; hence we can write
\begin{equation}
\frac{\left\langle \varphi_a^2 \right\rangle}{k_B T} = 2
\int^{\Lambda} \frac{d^2 k}{4 \pi^2} \frac{1}{k^2 + R} + F(
\mathcal{G}) \label{varphi2}
\end{equation}
where $F(\mathcal{G})$ is a universal function. A number of
analytic results for this universal function can be obtained from
the methods of Ref~\onlinecite{ssrelax}, and details appear in
Appendix~\ref{feyn}. For $\mathcal{G} \rightarrow 0$
(corresponding to $T \gg T_c$), perturbation theory in powers of
$U$ about the $\varphi_a =0$ saddle point yields
\begin{equation}
F( \mathcal{G}  \rightarrow 0) = (2.355711 \times 10^{-4})
\mathcal{G}^2 + \mathcal{O}(\mathcal{G}^3) \label{smallG}
\end{equation}
All subsequent terms in the above expansion involve only integer
powers of $\mathcal{G}$ and there are no logarithms. For
$\mathcal{G} \rightarrow \infty$ (corresponding to $T \ll T_c$),
we expand about a saddle point with $\varphi_a \neq 0$. As shown
in Ref.~\onlinecite{ssrelax}, this is done by introducing a `dual'
coupling $\mathcal{G}_D$ related to $\mathcal{G}$ by
\begin{equation}
\frac{1}{\mathcal{G}} + \frac{1}{2\mathcal{G}_D} = \frac{1}{6 \pi}
\ln \left( \frac{\mathcal{G}}{\mathcal{G}_D} \right).
\label{defgd}
\end{equation}
Note that as $\mathcal{G}\rightarrow \infty$, $\mathcal{G}_D = 3
\pi /\ln \mathcal{G}$. For large $\mathcal{G}$, the expansion of
$F$ is
\begin{equation}
F( \mathcal{G}  \rightarrow \infty) =  \frac{1}{2 \pi} \ln
\left(\frac{\mathcal{G}}{\mathcal{G}_D} \right) -
\frac{6}{\mathcal{G}} + \mathcal{O}(\mathcal{G}_D^2).
\label{largeG}
\end{equation}
All subsequent terms in the present expansion involve only integer
powers of $\mathcal{G}_D$, with no additional logarithms. As
discussed in Appendix~\ref{bosegas}, the numerical results of
Ref.~\onlinecite{nikolay1,nikolay2} yield the values of $F$ for
{\em all} values of $\mathcal{G}$. In particular, at the critical
point $\mathcal{G} = \mathcal{G}_c$ we have from (\ref{xiG})
\begin{equation}
F(\mathcal{G}_c) = 0.502 \pm 0.003 \label{FGc}
\end{equation}
The theory of the Kosterlitz-Thouless transition implies that
$F(\mathcal{G})$ will have a weak essential singularity at
$\mathcal{G} = \mathcal{G}_c$, similar to that in the specific
heat. A plot of the values of $F(\mathcal{G})$ appears in
Fig~\ref{plotF}. It is interesting to note that either the small
$\mathcal{G}$ or the small $\mathcal{G}_D$ expansions is accurate
for the entire range of $\mathcal{G}$ values.

The discussion so far presents our most complete results for the
properties of $\mathcal{F}_{GL}$ and $\mathcal{F}_{\varphi}$ with
essentially no approximations. There is, however, still some
residual cutoff dependence. This can be removed by subtracting
corresponding results at two different values of the bare coupling
$\widetilde{R}/T$ (or $m^{\ast} a(T)/T$), while $U$ fixed.
Depending upon the physical situation, changing $\widetilde{R}/T$
may also  involve some changes in the values of $U$. However, such
changes are expected to be small and we neglect any $T$ dependence
in $U$ and $\Lambda$ in the remainder of this section. This allows
to obtain an explicit relation between the dimensionless number
$\mathcal{G}$ used in the present section, and the number $g$ in
(\ref{defg}). Dividing (\ref{defR}) by $k_BT$ and subtracting the
corresponding equation at the critical point, and using the
definitions in (\ref{defall}) and (\ref{defG}), we obtain
\begin{equation}
g = \frac{6}{\mathcal{G}} - \frac{6}{\mathcal{G}_c} +
\frac{1}{\pi} \ln \left( \frac{T\mathcal{G}_c}{T_c\mathcal{G}}
\right). \label{defgG}
\end{equation}
As expected, $g$ extends from $+ \infty$ to $-\infty$ as
$\mathcal{G}$ extends from 0 to $+\infty$.
 Applying the same procedure to (\ref{varphi2}) we obtain the
 universal function in (\ref{asymp1})
\begin{equation}
\mathcal{D} (g, T/T_c)  = \frac{1}{2 \pi} \ln \left( \frac{T_c
\mathcal{G}}{T\mathcal{G}_c} \right) + F(\mathcal{G}) -
F(\mathcal{G}_c) \label{final}
\end{equation}
The expressions (\ref{Gc}), (\ref{smallG})-(\ref{final})
constitute the central results of this paper. Using as input the
values of $g$ and $T/T_c$, we compute $\mathcal{G}$ from
(\ref{defgG}) and $\mathcal{G}_D$ from (\ref{defgd}); then using
results (\ref{smallG}) and (\ref{largeG}) we can compute
$F(\mathcal{G}$), and finally insert in (\ref{final}) to obtain
$\mathcal{D}(g, T/T_c)$. In particular, the small $\mathcal{G}$
expansion in (\ref{smallG}) yields (\ref{asymp2}). Of course, it
is better to numerically solve for $\mathcal{G}$ from
(\ref{defgG}), rather than obtaining the solution order-by-order
in $1/g$ as was done for (\ref{asymp2}).

We now present some numerical results for the parameters used by
Mukerjee and Huse \cite{huse2}. They set $a(T)= a_0 (T -
T_c^{MF})$. Inserting this in (\ref{defg}) yields
\begin{equation}
g=  \frac{\rho_s (0)}{k_B T_c} \left(1 - \frac{T_c}{T}\right).
\label{mh}
\end{equation}
The parameterization $a(T) = a_0 (T-T_c^{MF})$ is chosen to be
valid near $T_c$, but can also be reasonably extended as $T
\rightarrow 0$ (in BCS theory, we expect a divergent $a(T
\rightarrow 0) \sim - \ln (1/T)$, but this divergence is expected
to be cutoff near a superfluid-insulator transition). By $\rho_s
(0) \equiv -\hbar^2 a(0) /(m^{\ast} b)$ in (\ref{mh}), we mean the
value of the helicity modulus of $\mathcal{Z}_{GL}$ extrapolated
to $T=0$ in this manner (the London penetration depth is related
to the helicity modulus by $\lambda_L^{-2} = 16 \pi e^{2} \rho_s
(T)/(\hbar^2 c^2)$). It is worth noting here that $\rho_s (0)$ and
$T_c$ are, in general, independent of each other, and the
Nelson-Kosterlitz relation \cite{nk} only constrains $\rho_s
(T_c)/T_c = 2/\pi$.

This framework now predicts all physical properties with 2 input
parameters: the values of $\rho_s (0)$ and $T_c$. Mukerjee and
Huse \cite{huse2} also defined a parameter $\eta$ as a measure of
the strength of fluctuations. This is related to the parameters
used here by $\eta = 2 k_B T_c^{MF}/\rho_s (0)$. In our numerical
results below, we set $\rho_s (0)/(k_B T_c) = (2/\eta)
(T_c^{MF}/T_c) = 6.8$, following their parameters.

An important subtlety should be noted here. The use of (\ref{mh})
in (\ref{defgG}) normally yields a value for $\mathcal{G}$ which
decreases monotonically with increasing $T$, as seems reasonable,
given our understanding of physical properties of the continuum
theory. However, because the value of $g$ in (\ref{mh}) saturates
as $T \rightarrow \infty$ and because of the presence of the
$\ln(T/T_c)$ term on the right-hand-side of (\ref{defgG}), for
very $T$ the value of $\mathcal{G}$ eventually starts increasing
with increasing $T$. This is clearly unphysical, and is an
indication that the present continuum theory breaks down at large
enough $T$. For the value of $\overline{\eta}$ being used here,
this unphysical non-monotonicity arises only at $T/T_c
> 20$, and we will therefore restrict our attention to values of
$T$ below this.

\begin{figure}
\centerline{\includegraphics[width=4.0in]{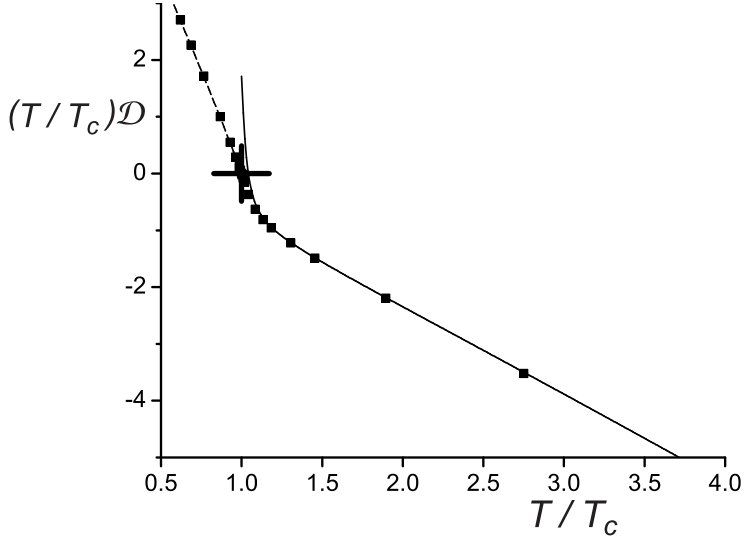}}
\caption{Plots of the universal function $(T/T_c) \mathcal{D} (g,
T/T_c)$ as a function of $T/T_c$ for $\rho_s (0)/(k_B T_c) = 6.8$.
>From (\ref{asymp1}) we see that $\langle | \Psi |^2 \rangle_T =
(T/T_c) \langle | \Psi |^2 \rangle_{T_c} + (m^{\ast} k_B
T_c/\hbar^2) (T/T_c) \mathcal{D} (g, T/T_c)$; so $\langle | \Psi
|^2 \rangle_T$ is determined from the above plot up to an
additive, non-singular, linear dependence on $T$ determined by
$\langle | \Psi |^2 \rangle_{T_c}$. This linear $T$ dependence can
compensate for the the linear $T$ dependence in the plot above so
that  $\langle | \Psi |^2 \rangle_T$ saturates at high $T$. Also,
as noted in the text, the present theory breaks down at large
enough $T$, and its main utility is in capturing the singular
increase in $\langle | \Psi |^2 \rangle_T$ as $T$ crosses $T_c$.
The solid line is the small $\mathcal{G}$ approximation obtained
by solving (\ref{Gc}), (\ref{smallG}), (\ref{FGc}), (\ref{defgG}),
(\ref{final}), and (\ref{mh}). The dashed line is the large
$\mathcal{G}$ approximation obtained by solving (\ref{Gc}),
(\ref{defgd}), (\ref{largeG}), (\ref{FGc}), (\ref{defgG}),
(\ref{final}), and (\ref{mh}). The square symbols are the
numerical data of Ref.~\protect\onlinecite{nikolay2} processed via
(\ref{defgG}), (\ref{final}), (\ref{mh}), (\ref{lambda}), and
(\ref{defX}). The plus marks the position of the Kosterlitz
Thouless transition.} \label{fig2}
\end{figure}

Solving (\ref{mh}) and (\ref{defgG}) for $\mathcal{G}$ as a
function of $T/T_c$, we use the results of this section and
Appendix~\ref{bosegas} to obtain the plot of Fig~\ref{fig2} for
the quantity appearing in (\ref{eq10}).
Note, again that either the small $\mathcal{G}$ or the large
$\mathcal{G}$ expansion is reasonable accurate.

\section{Conclusions}
\label{sec:conc}

We conclude this paper by discussing some of the experimental and
broader implications of our work.

Our primary result (\ref{eq10}) for the coherence length of the
charge order can be tested against neutron scattering and scanning
tunnelling spectroscopy (STS) experiments. However, the strong
random field disorder may make $\xi_{\phi}$ inaccessible to a
neutron probe which averages over the entire sample. In contrast,
STS provides a local probe, and so may be more sensitive to the
effects discussed here.

Consider an STS experiment with a field of view of area $A$, such
as those performed in
Refs.~\onlinecite{seamus,howald,krmp,seamus2,ali}. Quasiparticle
interference contributions, such as those computed in
Ref.~\onlinecite{dunghai,doug,tolyastm,franz2,yeh}, appear at low
temperatures, but we can expect that these will significantly
broaden at temperartures above $T_c$. We therefore focus here only
on the contribution of the $\phi$ fluctuations, which also lead to
modulations in the local density of the states measured in STS, as
shown in Ref.~\onlinecite{tolyastm,yeh}. We know that the STS
measurements are in the linear response regime. So, when we
perform the Fourier transform of the local density of states at
the ordering wavevector ${\bf K}_c$, we find that the signal is
proportional to the uniform part of the charge order parameter,
$\tilde{\phi} \equiv \phi ({\bf q}=0)$.  Let us estimate
$\tilde{\phi}$.  We have for the free energy
\begin{eqnarray}
\langle {\cal F} \rangle_{{\bf q}=0,A} \simeq -\langle h^2
\rangle^{1/2} \tilde{\phi} A^{1/2} +\xi_\phi^{-2} \tilde{\phi}^2 A
\label{FinA}
\end{eqnarray}
Here we used that the result that the random field energy scales
as the square root of the area. We can now minimize (\ref{FinA})
with respect to $\tilde{\phi}$
\begin{eqnarray}
\tilde{\phi} \sim \frac{\langle h^2 \rangle^{1/2}\xi_\phi^{2}
}{A^{1/2}}
\end{eqnarray}
Taking $\xi_\phi^{2}(T)$ from Eq. (\ref{eq10}) obtain get the
temperature dependence of the STS signal at the wavevector ${\bf
K}_c$.

For the case of competition between the superconducting ($\Psi$)
and charge ($\phi$) orders, the coupling $\lambda$ in (\ref{eq10})
will be positive. In this situation we have a seemingly
counterintuitive effect: as $T$ is increased through $T_c$, the
amplitude of the charge order is enhanced. The physical origin of
this is not difficult to understand: the increase in phase
coherence as $T$ is lowered is associated with an enhanced
coherent motion of the Cooper pairs, and this leads to a decrease
in the amplitude of the spatial modulations \cite{foot}.

An alternative statement of the same physics can be made in terms
of the vortices. As we argued in Ref.~\onlinecite{prl}, vortices
nucleate static charge order, and this was proposed as an
explanation of the experiments of Ref.~\onlinecite{seamus} (other
approaches\cite{so5,chenting,zhu,zhang,ghosal,andersen02} have
proposed static spin order in the vortices---in our theory, static
spin order is not nucleated by vortices and appears only in phases
with global magnetic order \cite{prl}). Increasing $T$ above $T_c$
causes a proliferation of vortices, and hence an enhancement of
charge order.

While our discussion in this paper has been entirely at the level
of the Landau theory of multiple order parameters, it is important
to keep in mind that such a theory is an effective model, and does
not preclude other interpretations which focus directly on the
electronic quasiparticles. In particular we can view the
competition between charge order and superconductivity as the
competition for the ordering of low energy quasiparticles near the
Fermi surface. So as the superconducting pairing of these
quasiparticles is reduced above $T_c$, they are more susceptible
to charge ordering.

An interesting direction for future work is to combine the
continuum theory of the Ginzburg-Landau functional presented here
with the theory of time-dependent superconducting fluctuations
presented in Refs.~\onlinecite{huse1,ussi,huse2}: this has the
prospect of placing more precise quantitative constraints on the
analysis of the Nernst effect experiments. Moreover, the accuracy
of either the small $\mathcal{G}$ or large $\mathcal{G}$
expansions suggests that useful analytic results may be possible.
Results for the fluctuation conductivity in such an approach,
including corrections to the Aslamazov-Larkin fluctuation
conductivity, have appeared recently \cite{sscond}. A similar
dynamic approach can also be applied to computing the linewidths
of the electronic quasiparticles in the pseudogap regime: the
strong amplitude fluctuations in $\Psi$ should lead to significant
broadening in the electronic spectral functions measured in
photoemission experiments.

\begin{acknowledgments}
We thank A.~Yazdani for numerous stimulating discussions in which
he proposed a related physical picture. We also thank J.~C.~Davis,
J.~Hoffman, A.~Kapitulnik, S.~Kivelson, D.~H.~Lee, D.~Podolsky,
N.~Prokof'ev, B.~Svistinov, N.-C. Yeh, A.~P.~Young, and
S.-C.~Zhang for useful discussions. This research was supported by
US NSF Grants DMR-0098226 (S.S.) and DMR-0132874 (E.D.) and the
Sloan Foundation.

\end{acknowledgments}

\appendix

\section{Correspondence with the dilute Bose gas}
\label{bosegas}

This appendix discusses the connection between the analysis of the
dilute Bose gas in Refs.~\onlinecite{nikolay1,nikolay2} and the
results of Ref.~\onlinecite{ssrelax} and the present paper. Let us
make it clear at the outset that we are not advocating a dilute
Bose gas description of the underdoped cuprates; rather, the
finite temperature properties of the dilute Bose gas are
characterized by some universal numbers which appear also in the
models of interest in the present paper.

The dilute Bose gas is defined by the partition function
\begin{eqnarray}
\mathcal{Z}_B &=& \int \mathcal{D} \psi ({\bf r}, \tau)
e^{-\mathcal{S}_B/\hbar} \nonumber \\
\mathcal{S}_B &=& \int_0^{\hbar /k_B T} d \tau \int d^2 r \left[
\hbar \psi^{\ast} \frac{\partial \psi}{\partial \tau} +
\frac{\hbar^2}{2m} \left| \nabla_{\bf r} \psi \right|^2 - \mu
|\psi |^2 + \frac{U_B}{2} |\psi |^4  \right].
\end{eqnarray}
We follow the notation of Refs.~\onlinecite{nikolay1,nikolay2}
throughout this appendix. The only exception is that the boson
interaction $U$ has been replaced by $U_B$, to prevent confusion
with the coupling $U$ in (\ref{fphi}).

Integrating out the non-zero Matsubara frequency modes in the Bose
gas, the action for the zero frequency modes takes the form
(\ref{fphi}) with the coupling constants
\begin{eqnarray}
- \frac{\hbar^2\widetilde{R}}{2m} &=& \mu - 2 U_B \int \frac{d^2
k}{4 \pi^2} \left( \frac{1}{e^{(\hbar^2 k^2 /(2 m) - \mu)/(k_B T)}
- 1}
- \frac{k_B T}{\hbar^2 k^2 /(2 m) - \mu} \right) \nonumber \\
U &=& \frac{12 m^2 U_B}{\hbar^4}
\end{eqnarray}
The integral above is divergent in the ultraviolet, but if we use
(\ref{defR}) to obtain the value of the renormalized coupling $R$
we obtain a convergent integral:
\begin{eqnarray}
R + \frac{2 m \mu}{\hbar^2} &=& \frac{4 m U_B}{\hbar^2} \int
\frac{d^2 k}{4 \pi^2} \left( \frac{1}{e^{(\hbar^2 k^2 /(2 m) -
\mu)/(k_B T)} - 1} - \frac{k_B T}{\hbar^2 k^2 /(2 m) -
\mu} + \frac{2 m k_B T/\hbar^2 }{k^2 + R}\right) \nonumber \\
&=& \frac{2 m^2 k_B T U_B}{\pi\hbar^4 } \ln \left( \frac{2 m
\mu}{\hbar^2 R (e^{\mu/(k_B T)}
- 1)} \right) \nonumber \\
& \approx & \frac{2 m^2 k_B T U_B}{\pi \hbar^4} \ln \left( \frac{2
m k_B T}{\hbar^2 R} \right). \label{eq11}
\end{eqnarray}
In the last expression we have expanded to leading order in $U_B$,
as required from consistency with previous approximations. Now
using the definition of the dimensionless coupling $\mathcal{G}$
in (\ref{defG}), we obtain the value of the chemical potential at
the Kosterlitz Thouless transition
\begin{equation}
\mu_c = \frac{m k_B T U_B}{\pi\hbar^2 } \ln \left[ \frac{\hbar^2
\xi_{\mu} }{m U_B} \right]
\end{equation}
with universal number $\xi_{\mu}$ computed in
Ref.~\onlinecite{nikolay1} related to the universal number
$\mathcal{G}_c$ computed earlier in Ref.~\onlinecite{ssrelax} by
\begin{equation}
\xi_{\mu} =  \frac{\mathcal{G}_c e^{- 6 \pi/\mathcal{G}_c}}{6}.
\label{ximG}
\end{equation}
Refs.~\onlinecite{nikolay1,nikolay2} obtained $\xi_{\mu}  = 13.3
\pm 0.4$, which is in reasonable agreement with the value
$\mathcal{G}_c \approx 102$ obtained in Ref.~\onlinecite{ssrelax};
the latter value of $\mathcal{G}_c$ yields $\xi_\mu \approx 14.1$
from (\ref{ximG}).

The same method can be used to compute the boson density $n$.
Integrating out the non-zero frequency modes and mapping onto the
classical theory (\ref{fphi}) we obtain
\begin{eqnarray}
n &=& \int \frac{d^2 k}{4 \pi^2} \left( \frac{1}{e^{(\hbar^2 k^2
/(2 m) - \mu)/T} - 1} - \frac{T}{\hbar^2 k^2 /(2 m) - \mu} \right)
+ \frac{m}{\hbar^2} \left\langle
\varphi_a^2 \right\rangle \nonumber \\
&=& \int \frac{d^2 k}{4 \pi^2} \left( \frac{1}{e^{(\hbar^2 k^2 /(2
m) - \mu)/T} - 1} - \frac{T}{\hbar^2 k^2 /(2 m) - \mu} + \frac{2 m
k_B T/\hbar^2 }{k^2 + R}
\right) + \frac{m k_B T}{\hbar^2} F (\mathcal{G}) \nonumber \\
&=& \frac{m k_B T}{2 \pi\hbar^2 } \ln \left( \frac{2m k_B
T}{\hbar^2 R} \right) + \frac{m k_B T}{\hbar^2} F (\mathcal{G});
\label{app1}
\end{eqnarray}
in the last equation we have made the same simplification as that
in the last equation in (\ref{eq11}). The result (\ref{app1})
yields the expression obtained in Ref.~\onlinecite{nikolay1} for
the critical density
\begin{equation}
n_c = \frac{m k_B T}{2 \pi\hbar^2} \ln \left( \frac{\hbar^2 \xi}{m
U_B} \right) \label{app2}
\end{equation}
with the universal number $\xi$ given by
\begin{equation}
\xi = \frac{\mathcal{G}_c}{6} e^{2 \pi F(\mathcal{G}_c)}.
\label{xiG}
\end{equation}
The simulations of Refs.~\onlinecite{nikolay1,nikolay2} obtained
$\xi = 380 \pm 3$, and inserting this result in (\ref{xiG}) allows
us to compute $F(\mathcal{G}_c )$.

Finally, subtracting (\ref{app2}) from the last equation in
(\ref{app1}) we obtain
\begin{eqnarray}
\frac{n-n_c}{m k_B T/\hbar^2} &=& \frac{1}{2 \pi} \ln \left(
\mathcal{G}/\mathcal{G}_c \right) + F(\mathcal{G}) -
F(\mathcal{G}_c) \nonumber \\
&=& \lambda (X). \label{lambda}
\end{eqnarray}
The function $\lambda (X) $ was computed in numerically
Ref.~\onlinecite{nikolay2}, and its argument $X$ can be related to
our coupling $\mathcal{G}$ by (\ref{eq11}), yielding
\begin{equation}
X = -\frac{6}{\mathcal{G}} + \frac{6}{\mathcal{G}_c} -
\frac{1}{\pi} \ln \left( \frac{\mathcal{G}_c}{\mathcal{G}}
\right). \label{defX}
\end{equation}
These earlier results for $\lambda (X)$ therefore yield the needed
function $F(\mathcal{G})$ from (\ref{lambda}) and (\ref{defX}).

\section{Weak and strong coupling expansions}
\label{feyn}

This appendix presents discusses the expansion for the universal
function $F(\mathcal{G})$ appearing in (\ref{varphi2}) for small
and large $\mathcal{G}$

\begin{figure}
\centerline{\includegraphics[width=4.0in]{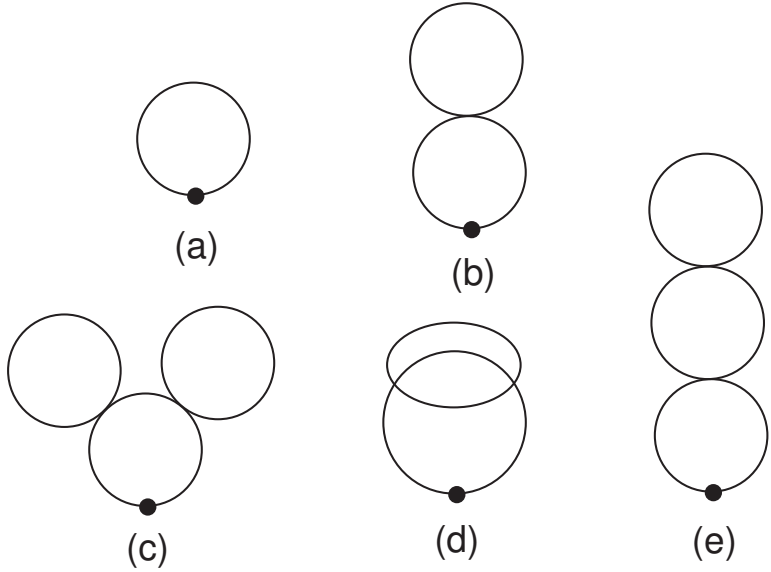}}
\caption{Feynman graph expansion of (\protect\ref{fphi}) for the
correlator (\protect\ref{varphi2}). } \label{fig3}
\end{figure}

For small $\mathcal{G}$, a simple Feynman graph expansion of
(\ref{fphi}) can be carried out to order $\mathcal{G}^2$, with the
diagrams shown in Fig~\ref{fig3}.
All propagators in Fig~\ref{fig3} involve the bare `mass'
$\widetilde{R}$. A simple calculation shows that the graphs
(a,b,c,e) are all absorbed by the first term on the
right-hand-side of (\ref{varphi2}), after substitution of
(\ref{defR}). This is a simple example of our claim that all
`tadpole' diagrams can be neglected after substituting $R$ for
$\widetilde{R}$. Only Fig~\ref{fig3}d contributes to $F
(\mathcal{G})$ in (\ref{varphi2}), and yields
\begin{eqnarray}
F(\mathcal{G}) &=& \frac{2\mathcal{G}^2}{9} \int \frac{d^2 p}{4
\pi^2} \int \frac{d^2 q}{4 \pi^2} \int \frac{d^2 k}{4 \pi^2}
\frac{1}{(k^2 + 1)^2 (q^2 + 1) ((p+k)^2 + 1) ((p+q)^2 + 1)}
\nonumber \\
&=& \frac{2\mathcal{G}^2}{9} \int \frac{d^2 p}{4 \pi^2} \left[
\frac{1}{2 \pi p \sqrt{p^2 + 4}} \ln \left( \frac{\sqrt{p^2 + 4} +
p}{\sqrt{p^2 + 4} - p} \right) \right] \nonumber \\
&~&~~~~~~~~~~~~~~~~~~~\times \left[ \frac{1}{4 \pi p (p^2 + 4)}
\left\{ p + \frac{2}{\sqrt{p^2 + 4}} \ln \left( \frac{\sqrt{p^2 +
4} + p}{\sqrt{p^2 + 4} - p} \right)\right\} \right]
\end{eqnarray}
We evaluated the last integral numerically and obtained
(\ref{smallG}).

For large $\mathcal{G}$, the method described in Appendix C of
Ref.~\onlinecite{ssrelax} was used. To one loop order, the result
\begin{equation}
F( \mathcal{G} ) = \frac{3 R_D}{k_B T U} + \frac{1}{2 \pi} \ln
\left( \frac{R}{R_D} \right) \label{frd}
\end{equation}
is easily obtained, where, as in Ref.~\onlinecite{ssrelax}, $R_D
\mathcal{G}_D = R \mathcal{G}$. In obtaining (\ref{frd}), we have
to explicitly account for all tadpole graphs, and the relationship
in (2.4) of Ref.~\onlinecite{ssrelax} between $\widetilde{R}$ and
$R_D$. The large momentum behavior of the expansion about
$\varphi_a = 0$ and $\varphi_a \neq 0$ saddle points should be the
same, and this ensures that the ultraviolet divergences cancel. At
two loop order, 35 Feynman graphs appear; these were evaluated as
in Appendix C of Ref.~\onlinecite{ssrelax}, and their sum was
found to vanish. Consequently, there is no order $\mathcal{G}_D$
term in $F$, and the result (\ref{largeG}) follows.


\begin{thebibliography}{}
\bibitem{3d} Three-dimensional effects are also important in a
narrow window of temperatures above $T_c$, where the transition
crosses over to the $d=3$ XY universality class. Such effects will
be neglected here.
\bibitem{gap} It is important to distinguish `amplitude
fluctuations' in the superconducting order from the `pairing gap'
of the electrons. The former measures the strength of the
superconducting correlations, while the latter is an energy gap
associated with the fermionic spectrum at high energies. In
particular, at low $T$ near a superconductor-insulator transition
it was argued in Refs~\protect\onlinecite{ssrelax,nature} that the
amplitude fluctuations become very strong. In contrast, the gap in
the electron spectrum remains finite across such a transition.
\bibitem{foot} However, it should be noted that the bare
coherence length $\xi_{\phi 0} (T)$ is likely to have a $T$
dependence which acts in the opposite direction: {\em i.e.} a
decrease in the amplitude of the charge order with increasing $T$.
This latter effect will not be singular near $T_c$, but should
eventually dominate at high enough $T$.
\end{thebibliography}
\end{document}